# The viscosity-radius relationship from scaling arguments

D. E. Dunstan

*Department of Chemical and Biomolecular Engineering, University of Melbourne, VIC 3010, Australia.*
davided@unimelb.edu.au

## Abstract

The viscosity-radius relationship for semi-dilute polymer solutions is derived using scaling arguments. The viscosity temperature and the temperature radius relationships are combined using the transitive equality relationship. The viscosity-radius relationship is determined to be $\eta \sim R^9$. The reduced viscosity observed with both increasing temperature and increasing shear rate then results from a reduction in coil radius.

A key assumption of polymer dynamics in flow is that the chains extend.[1-8] Here we present scaling arguments to show that chains compress in flow in a manner consistent with the observed decrease in viscosity of polymer solutions with temperature and the measured reduction of the radius with temperature.[3, 9-11]

Adam and Delsanti first presented the scaling argument to derive the viscosity-temperature relationship for semi-dilute polymer solutions [12]:

$$\eta \sim T^{\frac{-9(2-3\nu)}{3\nu-1}} \qquad [1]$$

For polymer solutions in a good solvent above critical overlap, the scaling exponents lie between 3/5 and 1/2.[13]

Assuming that the scaling exponent $\nu$, is 1/2 for the polymer in the concentrated regime yields:[4, 13-14]

$$\eta \sim T^{-9} \qquad [2]$$

Data presented in Bird et al. shows a viscosity temperature relationship: $\eta \sim T^{-17}$ that can be fitted using a scaling exponent of 4/9 showing the extreme sensitivity to the precise nature of the coil physical properties.[3] The temperature-radius relationship can be determined using the single chain Hookean force law derived from statistical mechanics.[4, 15-16] Assuming that the chains are ideal and that the entropic force determines the chain response to an external force yields the Hookean force law:

$$f = 3k_B Tr/R_0^2 \qquad [3]$$

Where f is the entropic force, $k_B$ Boltzmann's constant, T the absolute temperature r the average end to end vector of the chain and $R_0$ the end to end distance of the unperturbed chain. Here we note that the mean square of end-to-end distance of the chain is related to the radius of gyration such that $<r^2> = 6R_g^2$ where $R_g$ is the radius of gyration. Herein, we use R as the radius of the chain.

Equation 3 shows that for a given force, an increase in temperature results in a decreasing end to end distance of the chain. This behavior is generally observed for polymeric materials (elastomers) under tension. [9-10, 16-17] Physical measurements on polymeric materials show that the chains contract with increasing temperature as predicted by the models above.[4, 6, 11, 15, 18]

The "Theory of rubber elasticity" relates the shear modulus of the material, G, to the temperature[14, 16]:

$$G = NkT \qquad [4]$$

Where N is the density of entanglements. The definition of the modulus indicates that a concentrated polymer will contract as the temperature increases in a manner consistent with the single chain model (Equation 3) from which it is derived.[9, 11, 16, 19] It is assumed that no disentanglement occurs on the time scale of the measurement and that the number of entanglements, N, is therefore constant.

Furthermore, in a viscous flow experiment, energy is dissipated in the fluid in a manner proportional to the viscosity times the shear rate squared.[20] Increasing the thermal energy of the system results in chain contraction.[9, 11] As Mark correctly states Le Chatelier's principle, "since heat is given off during stretching, adding heat has to cause a contraction."[11] The viscosity of semi-dilute polymer solutions is observed to decrease with increasing temperature consistent with the chains contracting with temperature.[3, 21] We therefore use Equation 3 to derive the temperature radius relationship that is in accord with the experimental data.[9, 16] We assume that the form of Equation 3 is correct for a concentrated ensemble of chains.

Maintaining the system under constant stress while varying the temperature requires that the entangled system will show an inverse strain relationship with temperature. Assuming that the entangled system will deform affinely the strain will be proportional to the change in end to end distance.

Assuming that the force is approximately constant in Equation 3 then yields the relationship:

$$T \sim 1/R \qquad [5]$$

Combining [2] and [5] using the transitive property of equality yields[22]:

$$\eta \sim R^9 \qquad [6]$$

Equation 6 shows that the viscosity decreases with decreasing radius when the entropic contribution to the chain restoring force is considered. The decreasing viscosity observed in typical shear thinning then results from a decreasing coil size in solution.[2, 21] Equation 6 also shows an extreme sensitivity of the viscosity to changes in the coil size. Einstein suggested a viscosity-volume fraction squared relationship at high concentrations.[4, 7, 22] Equation 6 suggests a volume fraction cubed relationship. Scaling arguments predict that the viscosity follows a 14/3 power of the volume fraction at concentrations above the entanglement concentration.[7]

By assuming that the polymeric solutions show the empirically based power law behavior the radius-shear rate behavior may be predicted [1, 20]:

$$\eta \sim \dot{\gamma}^{-n} \qquad [7]$$

Where $\dot{\gamma}$ is the shear rate and the experimentally measured values for n typically lie between 1/2 and 1. [20, 23]

Combining [6] and [7] gives:

$$R \sim \dot{\gamma}^{-n/9} \qquad [8]$$

Recent experimental evidence is consistent with Equation 8 where the power of the shear rate radius relationship has been found to be 0.011 (~1/9) suggesting that the power law exponents are of order 1 for the polymers measured.[24-25] Furthermore, the model presented herein is validated by the experimental observation that polymer solutions reduce their viscosity with both increasing temperature and increasing shear rate.[2, 21] Both involve a reduction in the coil size. The model is also consistent with the observed extensional viscosities for these systems which show increasing viscosity with extensional shear indicating that R increases in these systems. We believe this to be of some importance to the field but leave this to others to determine.

## Acknowledgements.

I would like to acknowledge Maja Dunstan for proofing the manuscript. I would also like to thank the ARC for not funding this work. This gave me the time to think.